\documentclass[11pt]{article}
\usepackage{epsfig}
\usepackage[paper=a4paper,text={16cm,25cm}]{geometry}
\usepackage{multirow}
\usepackage{hhline}
\usepackage{amsmath}
\usepackage{amsfonts}
\usepackage{amssymb}

\usepackage[\ifnum\pdfoutput=1breaklinks\fi]{hyperref}

\newcommand{\Lower}[1]{\smash{\lower 1.5ex \hbox{#1}}}

\begin{document}
\title{New method to study ion-molecule reactions at low temperatures and application to the ${\rm H}_2^+ + {\rm H}_2\rightarrow {\rm H}_3^+ + {\rm H}$ reaction}
\author{P. Allmendinger, J. Deiglmayr, O. Schullian, K. H\"oveler, J. A. Agner, H. Schmutz, F. Merkt \\[3mm]
Laboratorium f\"ur Physikalische Chemie, ETH Z\"urich,
CH-8093 Zurich, Switzerland}
\date{\today}
\maketitle
\begin{abstract}
Studies of ion-molecule reactions at low temperatures are difficult because stray electric fields in the reaction volume affect the kinetic energy of charged reaction partners. We describe a new experimental approach to study ion-molecule reactions at low temperatures and present, as example, a measurement of the ${\rm H}_2^+ + {\rm H}_2\rightarrow {\rm H}_3^+ + {\rm H}$ reaction with the ${\rm H}_2^+$ ion prepared in a single rovibrational state at collision energies in the range $E_{\rm col}/k_{\rm B} = 5$-60 K. To reach such low collision energies, we use a merged-beam approach and observe the reaction within the orbit of a Rydberg electron, which shields the ions from stray fields. The first beam is a supersonic beam of pure ground-state H$_2$ molecules and the second is a supersonic beam of H$_2$ molecules excited to Rydberg-Stark states of principal quantum number $n$ selected in the range 20-40. Initially, the two beams propagate along axes separated by an angle of 10$^\circ$. To merge the two beams, the Rydberg molecules in the latter beam are deflected using a surface-electrode Rydberg-Stark deflector. The collision energies of the merged beams are determined by measuring the velocity distributions of the two beams and they are adjusted by changing the temperature of the pulsed valve used to generate the ground-state ${\rm H}_2$ beam and by adapting the electric-potential functions to the electrodes of the deflector. The collision energy is varied down to below $E_{\rm col}/k_{\rm B}= 10$~K, i.e., below $E_{\rm col}\approx 1$~meV, with an energy resolution of 100~$\mu$eV.  We demonstrate that the Rydberg electron acts as a spectator and does not affect the cross sections, which are found to closely follow a classical-Langevin-capture model in the collision-energy range investigated. Because all neutral atoms and molecules can be excited to Rydberg states, this method of studying ion-molecule reactions is applicable to reactions involving singly-charged cations.
\end{abstract}

\section{Introduction}

Ion-molecule reactions are very important chemical processes in low-density gaseous environments, which justifies the considerable attention they have received over the years. Many ion-molecule reactions have low reaction barriers or are barrierless and have large rate coefficients, even at very low temperatures. Consequently, they represent ideal systems with which to study chemical processes at temperatures close to 0 K, where one expects quantum-mechanical effects to become important (see, e.g., Refs.~\cite{althorpe-clary03a,nikitin-troe2005,iwm-smith06a,bell09a,gao11a} and references therein). Studies of ion-molecule reactions at low temperature are also important for the development of reliable models of astrochemical reaction cycles \cite{klemperer-herbst73,herbst03a,wakelam10,oka12a}.

The desire to study ion-molecule reactions at ever lower temperatures has stimulated a continuous stream of innovation for more than a century and numerous review articles provide a good overview of the progress achieved, experimentally and theoretically, in the understanding of low-temperature ion-molecule chemistry \cite{smith93,clary98,gerlich06}. Recently developed methods with which ion-molecule reactions can be studied at low temperatures include the use of uniform supersonic flows \cite{rowe89}, of multipole ion guides and buffer-gas cooling \cite{gerlich95,wester09}, and the use of ultracold ions in Coulomb crystals \cite{drewsen03,willitsch08,willitsch12}. A major difficulty encountered in experimental studies of ion-molecule reactions at the lowest temperature arises from stray electric fields. A variation of the electric potential of only 1 mV imparts 1 meV kinetic energy onto a singly-charged ion and thus hinders a control of the collision energy ($E_{\rm col}$) below this level, {\it i.e.}, below $E_{\rm col}/k_{\rm B}= 12$~K, where $k_{\rm B}$ is Boltzmann's constant. For this reason, even some of the most important ion-molecule reactions in the universe, such as the reaction ${\rm H}_2^+ + {\rm H}_2\rightarrow {\rm H}_3^+  +{\rm H}$, have not been studied experimentally so far at temperatures below about 50 K.

We present here an experimental approach to study ion-molecule reactions at collision energies below $E_{\rm col}= k_{\rm B}\cdot 10$~K which relies on the combination of the following three essential aspects: (i) Rather than studying an ion-molecule reaction in free space, we study it within the orbit of a Rydberg electron of high principal quantum number ($n>20$), which thus only acts as a distant spectator to the ion-molecule reaction. The advantage of this approach is that the reaction is shielded from electric stray fields by the Rydberg electron. That ion-molecule reactions can be studied in this way has been advocated and demonstrated by several authors \cite{pratt94,dehmer95a,wrede05,dai05,matsuzawa2010}.
(ii) To reach low collision energies, we merge two pulsed supersonic beams produced by pulsed valves capable of generating short pulses of about 10~$\mu$s duration. When the gas pulses propagate over large distances in a vacuum chamber, the molecules separate spatially according to their respective velocities and this enables the selection of molecules with very narrow velocity distributions. This advantage has been elegantly exploited in studies of reactions between neutral molecules at very low temperatures \cite{henson12,lavert-ofir14,jankunas14,bertsche14,jankunas15}. (iii) To merge a beam of Rydberg molecules with a beam of  ground-state molecules, we use an on-chip Rydberg-Stark surface-electrode deflector \cite{allmendinger14a}, which also serves the purpose of selecting a very restricted phase-space volume and thereby enhances the collision-energy resolution of the experiment.

This approach is in principle applicable to a broad range of ion-molecule reactions because all atoms and molecules can be excited to Rydberg states.  To illustrate it, we present a measurement of the relative integral cross section of the ${\rm H}_2^* + {\rm H}_2\rightarrow {\rm H}_3^+ + {\rm H}+{\rm e}^-$ reaction (${\rm H}_2^*$ designates an H$_2$ molecule excited to a Rydberg state), which was shown to be equivalent to the reaction ${\rm H}_2^+ + {\rm H}_2\rightarrow {\rm H}_3^+ + {\rm H}$ \cite{pratt94}.  This reaction is very important in astrochemistry and has been extensively studied experimentally \cite{anderson81,shao86,pollard91,mackenzie94,pratt94} and theoretically~\cite{eaker1985,baer90}, in the low temperature regime most recently by Dashevskaya {\it et al.} \cite{dashevskaya05} and Sanz-Sanz {\it et al.} \cite{sanz-sanz15}. In their efforts to measure the integral cross section for this reaction at low collision energies, Glenewinkel-Meyer and Gerlich have been able to obtain reliable values down to $E_{\rm col}/k_{\rm B}= 60$~K~\cite{glenewinkel-meyer97}. Their data, with which the most recent theoretical results are compared \cite{sanz-sanz15}, still represent the experimental reference on this reaction at low collision energies. The results we present here extend the collision-energy range from the lowest collision energies reported by Glenewinkel-Meyer and Gerlich to $E_{\rm col}/k_{\rm B}= 5$~K, with selected measurements below 2~K.

\section{Experiment}

\subsection{Experimental setup and procedure}
\label{experiment}

A schematic view of the experimental setup is presented in Fig.~\ref{fig1}. It consists of differentially pumped beam-source chambers and a reaction chamber with a time-of-flight mass spectrometer. The beam-source chambers are separated from the reaction chamber by a 6-cm-long region containing a Rydberg-Stark surface-electrode deflector.
Two pulsed collimated supersonic beams of pure H$_2$ are formed in the beam-source chambers using liquid-nitrogen-cooled pulsed valves and initially propagate at an angle of 10$^\circ$. After passing a skimmer, the two beams enter a high-vacuum chamber and are further collimated just before the reaction zone (see Fig.~\ref{fig1}). Just before the two beams cross, the molecules in one of these two beams are excited from the $\mathrm{X}\, ^1\Sigma_{\mathrm g}^+ \, (v=0,J=0)$ ground state to long-lived $nkm$ Rydberg-Stark states with principal quantum number $n$ in the range 20-40 and the ion core in the $\mathrm{X}^+\, ^2\Sigma_{\mathrm g}^+\,(v^+=0,N^+=0)$ state using the resonant three-photon excitation sequence
\begin{equation} \label{eq1}
nkm [\rm{X}^+\, ^2\Sigma_{\rm g}^+\,(0,0)] \leftarrow \rm{I}\, ^1\Pi_{\rm g} (0,2) \leftarrow \rm{B}\, ^1\Sigma_{\rm u}^+ \, (3,1) \leftarrow \rm{X}\, ^1\Sigma_{\rm g}^+ \, (0,0).
\end{equation}
This excitation sequence and the three lasers used for the successive excitation steps have been described in Refs.~\cite{hogan09a,seiler11b}. In the following, we refer to the beam containing  Rydberg H$_2$ molecules as the Rydberg H$_2$ beam and to the other beam as the ground-state H$_2$ beam, which consists of a statistical mixture of para and ortho H$_2$ molecules in their X $^1\Sigma_g^+\ (v=0,N=0)$ (25\%) and $(v=0,N=1)$ (75\%) ground state.

A surface-electrode deflector, consisting of a set of electrodes on a bent printed circuit board (PCB) (see Ref.~\cite{allmendinger14a} for details), is used to deflect the Rydberg molecules and to merge the two supersonic beams. A shutter labeled C in Fig.~\ref{fig1} prevents undeflected Rydberg molecules from entering the reaction zone. The deflector can also be used to select molecules with a narrower velocity distribution than the initial beam and to accelerate or decelerate them.

The relative central velocity of the two beams is set either by adjusting the temperature of the pulsed valve generating the ground-state H$_2$ beam while keeping the velocity of the Rydberg H$_2$ beam constant, or by adjusting the waveforms of the electric potentials applied to the electrodes of the deflector. Because a precise knowledge of the velocity distributions of the two beams is crucial for the determination of the energy dependence of the cross sections, they are measured independently. The velocity distribution of the ground-state H$_2$ beam is determined by photoexciting the H$_2$ molecules to long-lived Rydberg states at the spot labeled A in Fig.~\ref{fig1}. The photoexcitation is carried out before the molecules enter into the reaction zone using the same excitation sequence (Eq.~(\ref{eq1})) as used to generate the Rydberg molecules in the second beam. The long-lived Rydberg molecules then fly as neutral particles until they reach the microchannel-plate (MCP) detector labeled "MCP1" in Fig.~\ref{fig1}  and located beyond the reaction zone. The velocity distribution of the molecules excited at spot A during the 5-ns-long laser-excitation pulse is directly determined from the distribution of flight times to the detector. The full velocity distribution of the beam is reconstructed from measurements in which the delay between the valve-opening time and the laser-photoexcitation pulse is systematically varied over the entire duration of the gas pulse (see Fig.~\ref{fig2} and Section~\ref{velocity}).

The three-dimensional velocity distribution and the main geometrical parameters of the experimental configurations (e.g., the valve-opening time and the distances between the valve front plate, the excitation spot, the reaction zone and the detectors) are used as input parameters for the simulation of the trajectories of the ground-state H$_2$ and the Rydberg H$_2$ molecules in the vacuum chambers, as described in Section~\ref{simulations}. The procedure used to determine the velocity and density distributions of the deflected Rydberg H$_2$ beam has been described in Ref.~\cite{allmendinger14a}, to which we refer the reader for details (see also Section~\ref{simulations}).

The merged beams enter a reaction zone located within a cylindrical stack made of three electrodes, with which the reaction products are extracted towards an imaging detector (MCP with phosphor screen ("MCP2") and CCD camera) in a direction perpendicular to the merged-beam-propagation axis. The extraction electrodes have a diameter of 7~cm and are designed so that pulsed extraction fields can be applied in a Wiley-McLaren configuration \cite{wiley55a} to extract the ions and detect them mass selectively. The electric-field distribution generated in the reaction zone upon application of a potential difference across the stack is determined using a commercial simulation program (SIMION 8.1~\cite{simion}). The same program is also used to simulate the trajectories of the H$_3^+$ ions produced by the reaction and to calculate their time-of-flight distribution.

An essential aspect of the experiments presented in this article is linked to the ability of generating short pulses of cold molecules in well-collimated supersonic beams. The techniques required to generate short-pulse molecular beams are well established~\cite{gentry78a,even00a}. For experiments in which the velocity of a supersonic beam is tuned by varying the valve temperature, solenoid valves offer distinct advantages because they can be operated at cryogenic temperatures (see, e.g., Refs.~\cite{hagena91a,pentlehner09a}). To achieve pulse lengths of about 10~$\mu$s for the experiments presented in this article, special attention was paid to the miniaturization of the design following general principles established in the development of earlier valves \cite{gentry78a,hagena91a,even00a}.
Our home-built pulsed valves consist of a 1.5-mm-diameter cylindrical plunger (Vacuumschmelze Vacoflux50) located inside a ceramic tube. The plunger is spring-preloaded with a coil spring towards a 0.125-mm-thick polyimide disk sealing the front plate with a central orifice of 0.35~mm diameter. The plunger is actuated with a coil (0.4-mm-diameter copper wire, 65 windings). No specific measure is taken to guide or enhance the magnetic fields near the plunger. Instead, high pulsed currents of about 100 A are applied to the coil. To avoid eddy currents, the coil is surrounded by Al$_2$O$_3$ ceramic spacers. With this design, valve opening times as short as 13~$\mu$s could be achieved and were used in the present work.
The valve assembly is fully encapsulated in a copper body with epoxy (Stycast 2850ft). The pulsed valve is linked to a liquid-nitrogen Dewar using copper braids  and the valve temperature is regulated by heating.

Another essential aspect of the measurement procedure is that it is accompanied by numerical simulations of the reaction from the preparation of the molecules in the supersonic beams prior to the reaction to the detection of the product ions. The simulation procedure is outlined in Section~\ref{simulations}.

\subsection{Determination of velocity distributions and collision energies}
\label{velocity}

\begin{center}
\begin{figure}[ht!]
\includegraphics[width=15 cm]{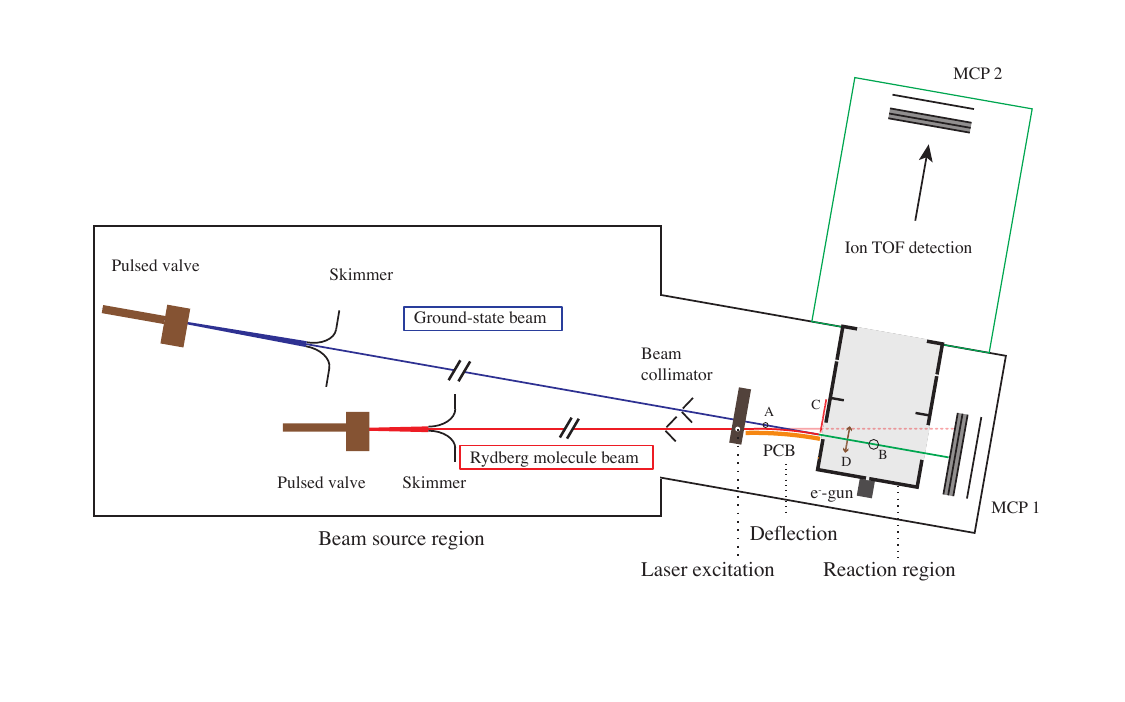}
\caption{Schematic representation of the merged-beam apparatus used to study ion-molecule reactions at low collision energies, with the two skimmed supersonic beams initially propagating at an angle of 10$^\circ$, the Rydberg-Stark deflector made of a bent printed circuit board (PCB) and used to merge the beams after laser excitation, the reaction zone, and the linear time-of-flight spectrometer used to detect reactants and products separately. (MCP1) and (MCP2) Microchannel-plate detectors to monitor the flight times of Rydberg H$_2$ molecules and the ion-time-of-flight spectra, respectively. (A) Laser excitation spot used in measurements of the velocity distribution of the undeflected beam of ground-state H$_2$ molecules. (B) Region in which the density profile of the ground-state molecules in the undeflected beam is measured by electron-impact ionization induced by an electron gun. (C) Aperture used to prevent the undeflected Rydberg molecules from entering the reaction region. (D) Movable microchannel-plate detector used for measurements of the velocity distribution and the lifetimes of the Rydberg molecules.}
\label{fig1}
\end{figure}
\end{center}

\begin{center}
\begin{figure}[ht!]
\includegraphics[width=15 cm]{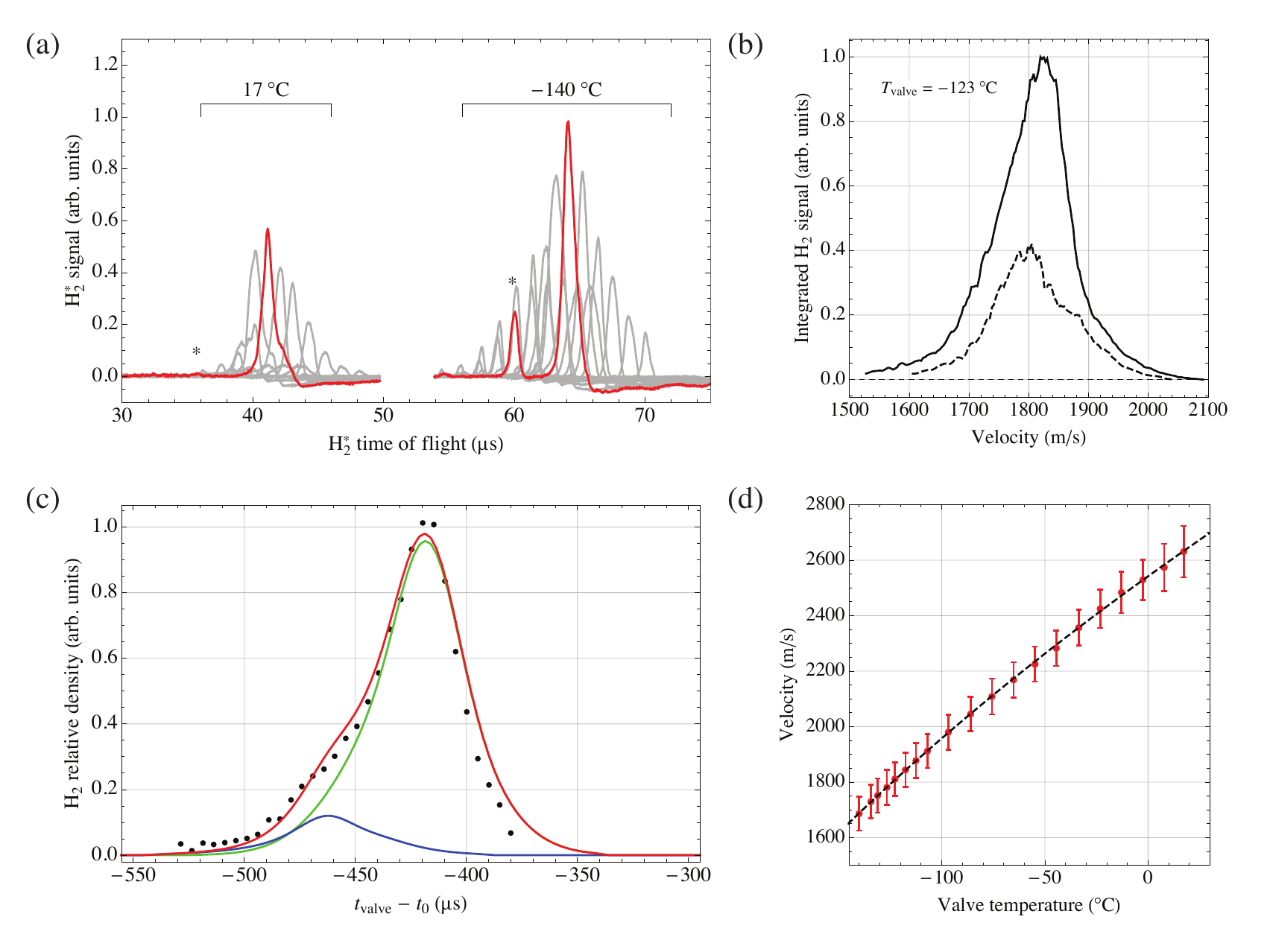}
\caption{Determination of the density and velocity distributions of H$_2$ molecules in the ground-state beam.
(a) Time-of-flight profiles of long-lived H$_2$ Rydberg molecules prepared by photoexcitation at the position marked A in Fig.~\ref{fig1} at different time delays between the valve opening and the photoexcitation laser pulse and for valve temperatures of $17~^\circ$C (left) and $-140~^\circ$ C (right). The red trace highlights the measurement carried out at the center of the gas pulse and the asterisk indicates the contribution from the plunger-rebounce pulse.
(b) Velocity distribution extracted for the main pulse (full line) and the plunger rebounce pulse (dashed line) from time-of-flight measurements such as those depicted in (a) but for a valve temperature of $-123~^\circ$C.
(c) Dots: Relative H$_2$ density in the ground-state beam measured by electron-impact ionization in the region labeled B in Fig.~\ref{fig1}. Red line: Particle trajectory simulation of the beam density based on the velocity distribution presented in panel (b) and decomposed into contributions from the main pulse (green line) and the pulse originating from the plunger rebounce (blue line). (d) Mean velocity (dots) and full width at half maximum of the velocity distribution (vertical bars) as a function of the valve temperature.
}
\label{fig2}
\end{figure}
\end{center}

The measurement of the velocity distribution of the H$_2$ molecules in the ground-state beam is carried out by photoexcitation to long-lived Rydberg states at the position marked A in Fig.~\ref{fig1} and recording their times of flight to the MCP detector labeled "MCP1" in Fig.~\ref{fig1}. Because of the short duration of the valve opening, the length of the gas pulse at the photoexcitation spot is almost entirely determined by the velocity distribution of the molecules in the supersonic beam and by the experimental geometry. The distances between the valve orifice and the skimmer and the photoexcitation spot are 7~and 80~cm, respectively. The fastest molecules in the beam arrive at the photoexcitation spot first and are excited at the shortest delay times between the opening of the pulsed valve and the laser pulse. Consequently, their flight times from the photoexcitation spot until ionization at the detector "MCP1" are shorter than the flight times of molecules photoexcited at longer delays. This behavior is illustrated by the time-of-flight distributions presented in panel (a) of Fig.~\ref{fig2}. When operating the valve at a temperature of 17$^\circ$C (left part of the panel), the H$_2$ molecules in the supersonic beam have velocities around 2600~m/s. The molecules excited by the lasers at the shortest (longest) delays arrive at the detector first (last) and their flight times from the excitation spot to MCP1 are  $\approx 40\ \mu$s (46~$\mu$s), corresponding to a velocity of $\approx 2780$~m/s (2420~m/s).

Cooling the valve to $-140^\circ$C reduces the speed of the supersonic expansion to about 1700~m/s, as illustrated by the time-of-flight profiles displayed on the right-hand side of Fig.~\ref{fig2}a.
These measurements demonstrate that pulsed laser excitation enables one to select molecules with a much narrower velocity distribution than the overall beam because of the dispersion taking place as the molecules travel from the valve orifice to the photoexcitation spot. This dispersion also naturally selects the velocity of the molecules present in the reaction zone at a given time. This selection is essential to achieve a high energy resolution in the measurement of reaction cross sections.

The time-of-flight traces presented in Fig.~\ref{fig2}a reveal that the velocity distributions are not uniform, but consist, for the short pulses generated by our valve, of two components. Indeed, when pulsing high currents to achieve an intense gas flow through the orifice of the valve, bouncing of the plunger is difficult to prevent~\cite{christen13a}. This second component is emitted by the valve at a later time. Consequently, the molecules released in this second pulse that are probed by the photoexcitation pulse have a higher velocity than the molecules originating from the main pulse. This behavior is clearly observable in the traces displayed in Fig.~\ref{fig2}a, as highlighted by the red trace, where the asterisk designates the contribution from the second gas pulse. Fortunately, the second component is well separated in time from the first and is only minor, so that its contribution to the reaction can either be eliminated by a careful adjustment of the timing of the experiment or precisely accounted for in the analysis.

The overall velocity distribution of the H$_2$ molecules in the supersonic beam extracted from time-of-flight measurements such as those presented in panel (a) is depicted in Fig.~\ref{fig2}b for a valve temperature of $-123^\circ$C. The velocity distributions follow in a straightforward manner from the known distance between laser-excitation spot and MCP surface. The velocity distribution is centered around 1820~m/s with a full width at half maximum of about 120~m/s. The dispersion taking place between the valve orifice and the reaction zone, however, strongly reduces the velocity distribution of the molecules contributing to the reaction. To model this dispersion and precisely characterize the velocity of the molecules reaching the reaction zone at a given time, we use the particle-trajectory-simulation program described in Section~\ref{simulations}. The velocity distributions obtained from measurements such as those presented in panels (a) and (b) are used as input distributions to the simulations of the ground-state H$_2$ beam and to extract the mean velocity of the main component of the gas pulse as a function of the temperature of the valve (see Fig.~\ref{fig2}d).

To validate these distributions, we carried out an independent measurement of the relative density of the H$_2$ molecules in the beam, and monitored the H$_2^+$ ion signal generated in the region labeled B in Fig.~\ref{fig1} by electron-impact ionization as a function of the trigger time and the temperature of the valve.
The ionization signal was monitored after extraction of the H$_2^+$ ions toward the MCP detector labeled "MCP2" in Fig.~\ref{fig1} and located at the end of the flight tube.
The results of a typical measurement are displayed as dots in Fig.~\ref{fig2}c, where they are compared to the results of a numerical simulation obtained with our particle-trajectory-simulation program (red trace). The simulations and their good agreement with the experimental results enable us to decompose the intensity distribution into the two components of the gas pulse, which are displayed as green and blue traces for the main component of the gas pulse and the minor component originating from the plunger rebounce, respectively. This quantification enables us to accurately predict the relative density and the velocities of the H$_2$ ground-state molecules present at a given time and a given position in the reaction zone and for a given valve temperature, which is crucial to determine reliable relative reaction cross sections. The combination of simulations and experiment indicates that the spatial extent of the ground-state-H$_2$-molecule packets along the beam-propagation direction is about 10~cm in the reaction zone. The ground-state-H$_2$-molecule packets are thus longer than the diameter (7~cm) of the ion-extraction electrodes.

The velocity distribution of the Rydberg H$_2$ beam at the exit of the Rydberg-Stark deflector can also be precisely determined, as explained in Ref.~\cite{allmendinger14a}. The spatial length of the H$_2$-Rydberg-molecule packet along the beam-propagation direction in the reaction zone is only a few millimeters and thus much shorter than the length of the packet of ground-state H$_2$ molecules because (1) photoexcitation at the entrance of the deflector selects a narrow slice of the original velocity distribution, (2) the deflector is operated in a guiding mode with a small trap volume of $\approx 1$~mm$^3$ and preserves the velocity component in the instantaneous propagation direction of the beam, with minimal heating to at most 300 mK~\cite{allmendinger14a}, and (3) the reaction zone is located immediately after the deflector (see Fig.~\ref{fig1}).

The distribution of collision energies between the ground-state and the Rydberg H$_2$ molecules is determined by the spatial overlap of the short H$_2$-Rydberg-molecule packet with the long ground-state H$_2$ packet in the time interval during which the reaction is monitored. As time proceeds, the H$_2$ Rydberg molecules interact with ground-state H$_2$ molecules of changing average velocity. Consequently, the average collision energy changes with time as the H$_2$-Rydberg-molecule packet moves across the ground-state-H$_2$-molecule packet. This effect reduces the collision-energy resolution of the measurements.

Because of the large average kinetic energy of the H$_3^+$ reaction products (about 0.17 eV, corresponding to an average velocity of about 3500 m/s in the center-of-mass frame~\cite{pollard91}), most of the H$_3^+$ product ions leave the reaction zone within 10~$\mu$s of their formation. The probability that an H$_3^+$ product ion is still in the reaction zone at a given time after the reaction decreases with time and can be estimated from the kinetic-energy distribution of the reaction products (see Section~\ref{simulations}). The collision-energy resolution thus initially deteriorates with increasing interaction time
and saturates to a value of about $k_{\rm B}\cdot 1$~K when the interaction time approaches 10~$\mu$s. This effect can be accurately modelled from a known or assumed product-kinetic-energy distribution using our particle-trajectory-simulation program.

The time interval during which the reaction is monitored can be restricted experimentally by applying two pulsed electric fields across the stack of extraction electrodes surrounding the reaction zone (see Fig.~\ref{fig1}). The first pulse sweeps all ions out of the reaction zone and initializes the reaction-observation time, which lasts until the application of the second pulse. This second pulse exclusively extracts the H$_3^+$ ions produced by reactions having taken place during the field-free interval between the two pulses. It also field ionizes the H$_2$ Rydberg molecules that have not reacted provided that their principal quantum number is high enough for field ionization. For all experiments presented below, we chose a value of 26~V/cm for both pulses, which is insufficient to field ionize the Rydberg states in the range of $n$ values between 20 and 40 prepared optically, but efficiently field ionizes Rydberg states with $n$ beyond 60 produced by blackbody-radiation-induced transitions as the H$_2$ Rydberg molecules propagate from the entrance of the deflector to the reaction zone~\cite{seiler16a}. When adjusting the length of the interval between the two field pulses, a compromise must be made. Indeed, too short an interval reduces the number of H$_3^+$ ions whereas too long an interval deteriorates the collision-energy resolution of the measurements for the reasons given above.
With an interval of 3~$\mu$s, the collision-energy resolution is limited to $k_{\rm B}\cdot 300$~mK by the temperature of the Rydberg H$_2$ beam.

\section{Particle-trajectory simulation of collision experiments}
\label{simulations}
In this section, we describe in detail the simulations of the H$_3^+$-product-ion yield of the H$_2^+$ + H$_2$ reaction. The simulations are based on
\begin{enumerate}
\item \label{enum:trajectories} trajectories $\vec{r}(t)$ of deflected Rydberg H$_2$ molecules. The trajectories are obtained by classical, numerical propagation of particles  with randomly sampled initial conditions as described in Ref.~\cite{allmendinger14a}.
\item \label{enum:neutralh2} the density distribution $\rho_{\mathrm{H}_2}(\vec{r},t)$ and the velocity distribution $\vec{v}_{\mathrm{H}_2}(\vec{r},t)$ of the ground-state H$_2$ molecules. The distributions are obtained by analytic propagation of measured velocity and density profiles as described in Section~\ref{determinationOfVelocitySim}.
\item \label{enum:reaction} a pseudo-first-order rate model for the reaction of isolated Rydberg H$_2$ molecules with the much more abundant ground-state H$_2$ molecules under conditions where the total reaction probability per Rydberg molecule is much less than one.
\item \label{enum:crosssections} the differential cross sections as a function of the recoil energy and the scattering angle based on previous experimental measurements~\cite{pollard91}.
\item \label{enum:detection}  the detection probability of the product ions resulting from classical trajectory simulations of the time-of-flight detection process. The simulation of the reaction and detection process is implemented numerically by the Monte-Carlo method.
\end{enumerate}

To illustrate the general concept of the simulations, we describe in the following the fate of a fictive Rydberg H$_2$ molecule created in the moving trap above the chip-based deflector. The trajectory of this particle (item \ref{enum:trajectories} in the list presented above), which is determined purely by classical mechanics, is calculated by solving Newton's equation of motion in the applied inhomogeneous electric fields during deflection, and in free space once the particle is released from the trap. The accuracy of calculations of this type has been verified in numerous experimental studies~\cite{ seiler11b,allmendinger14a}. In the next step, the Monte-Carlo simulation of the reaction process (item \ref{enum:reaction}) samples randomly the probability for a reaction to occur along the Rydberg trajectory
\begin{equation}
P_{\mathrm{reaction}}(t,t+\Delta t)=k_{\mathrm{eff}} \, \rho_{\mathrm{H}_2}(\vec{r}(t),t)  \Delta t \; ,
\label{rate}
\end{equation}
where $k_{\mathrm{eff}}$ is an effective reaction-rate coefficient which is assumed to be constant over the range of collision energies probed along the trajectory of the H$_2$ Rydberg molecule. Random sampling of equation~(\ref{rate}) faithfully reproduces the distribution of product ions if the total reaction probability per Rydberg molecule is much less than one, \textit{i.e.}, if the collision-free path length is much larger than the dimensions of the interaction region. Experimentally this condition is fulfilled because there is no measurable reduction of the number of Rydberg molecules, detected after the reaction zone, when the beam of ground-state H$_2$ molecules is introduced. The velocity vectors of the product ions are randomly sampled from the assumed differential cross sections (item \ref{enum:crosssections}). The classical motion of the product ions before and during extraction in the time-of-flight stack (see Fig.~\ref{fig1}) is simulated, again by solving Newton's equation of motion in time-dependent electric fields (item \ref{enum:detection}). The number of simulated product ions which reach the detector (MCP2 in Fig.~\ref{fig1}) within the experimental time-of-flight window is counted.

The number of counted product ions, averaged over many Rydberg-molecule trajectories, is the result of the simulation to which the experimental measurements are compared. The simulations are expected to reproduce the experimental observations only up to factor, which depends on the number of ground-state H$_2$ molecules in the pulse and the detection efficiency of the MCP. This factor does not, however, depend on the collision energy, so that any energy-dependent variation of the ratio between the simulated and the measured number of product ions can be directly related to an energy-dependence of $k_{\mathrm{eff}}$.

In the following, we describe in more detail the determination of the density and velocity distributions of the ground-state H$_2$ molecules from measured time-of-flight distributions (item \ref{enum:neutralh2}), and the distribution of product-ion-velocity vectors on the basis of differential cross sections (item \ref{enum:crosssections}).

\subsection{Velocity and density distributions of the ground-state H$_2$ molecules}
\label{determinationOfVelocitySim}
The determination of density and velocity distributions of the ground-state H$_2$ molecules is based on the laser excitation of H$_2$ molecules to a high-$n$-Rydberg state at a well-defined spot with $z = z_{\mathrm{exc}}$ (labeled A in Fig.~\ref{fig1}) and time $t_{\mathrm{exc}}$, as described in Section~\ref{velocity}. The measured time of flight $t_{\mathrm{TOF}}$ of the molecules from $z_{\mathrm{exc}}$ to the front of MCP1 at position $z_{\mathrm{MCP1}}$ (see Fig.~\ref{fig1}) yields directly the velocity of the excited molecules in the beam direction (chosen as the $z$ axis in the following). For a given temperature of the valve $T_{\mathrm{valve}}$, $t_{\mathrm{TOF}}$ depends only on the difference $\Delta t  = t_{\mathrm{exc}}-t_{\mathrm{valve}} $ between the valve-opening time $t_{\mathrm{valve}}$ and $t_{\mathrm{exc}}$ and one finds
\begin{equation}\label{eq:sim:measuredv}
v(\Delta t) =  \frac{z_{\mathrm{MCP1}}-z_{\mathrm{exc}}}
{t_{\mathrm{TOF}}(\Delta t)},\;
\end{equation}
where we have omitted the implicit dependence on $T_{\mathrm{valve}}$. To simulate a scattering experiment performed for a fixed value of $t_{\mathrm{valve}}$, the ground-state H$_2$ velocity needs to be known at arbitrary positions $(\vec{r},t)$. We therefore propagate $v(\Delta t_i)$, measured at discrete values $\Delta t_i$ of the excitation delay, to new positions
\begin{equation}\label{eq:sim:newpos}
z_i(t) = z_{\mathrm{exc}} + v(\Delta t_i)\left(t-\Delta t_i-t_{\mathrm{valve}}\right) \; .
\end{equation}
Because the measured velocity distribution $v(\Delta t)$ is a monotonously-decreasing function, this procedure yields uniquely defined values $v(z_i(t),t)$, which are interpolated in $z$ to obtain $v(z,t)$. The radial $x$ and $y$ components of the velocity vectors are determined geometrically, assuming propagation along a straight line with origin at the valve orifice.

The number of Rydberg molecules detected on MCP1 after laser excitation at a delay $\Delta t$ is proportional to the number $\mathcal{N}(z_{\mathrm{exc}},\Delta t)$ of ground-state molecules in the laser-excitation volume at $z = z_{\mathrm{exc}}$. For the simulation of the reaction (item \ref{enum:reaction}), the density of ground-state molecules is required as a function of arbitrary coordinates $\vec{r}$ and times $t$. To this end, we first transform the distribution $\mathcal{N}(z_{\mathrm{exc}},\Delta t)$, measured in the time domain at a fixed position, into a spatial distribution at a fixed time. This is done numerically using the procedure described above for the determination of $v(z,t)$ by propagating $\mathcal{N}(z_{\mathrm{exc}},\Delta t)$ to new positions $z_i$ according to
\begin{equation}\label{H2density}
\tilde{N}(z_i(t),t) =  \mathcal{N}(z_{\mathrm{exc}},\Delta t_i) \left(\frac{\mathrm{d} \Delta t}{\mathrm{d} z}\right)_i,
\end{equation}
where the derivative, which is evaluated numerically, ensures the conservation of the particle number $N_\mathrm{exc} = \int \tilde{N}(z , t) d z$ at all times $t$.

The relative density of ground-state H$_2$ molecules along $z$ at time $t$ is then given by
 \begin{equation}
 \tilde{\rho}(z_i,t) =  \frac{1}{N_\mathrm{exc}}
 \frac{1}{(z_i-z_{\mathrm{valve}})^2} \tilde{N}(z_i, t)  \; .
\end{equation}
The factor $1/(z-z_{\mathrm{valve}})^2$ accounts for the radial expansion of the ground-state H$_2$ pulse as it propagates away from the point-like valve orifice.
We then obtain $\tilde{\rho}(z,t)$, again by interpolation. The quantity $\tilde{\rho}(z,t)$, which is determined separately at many different temperatures of the valve, does not depend any more on experimental parameters such as the excitation volume, the absolute density of ground-state molecules, or the probability for exciting a ground-state molecule into a long-lived Rydberg state. To account for the variation of the absolute ground-state-molecule density with $T_{\mathrm{valve}}$, we multiply $\tilde{\rho}(z,t)$ by a factor which is proportional to the total number of ground-state molecules in the gas pulse as measured by electron-impact ionization (see Section~\ref{experiment}) and obtain $\rho_{{{\mathrm{H}}_2}}^{T_{\mathrm{valve}}}(z,t)$. In principle, the proportionality constant could be determined from the efficiency of the electron-impact-ionization process and the gain factor of the MCP detector. In this work, however, we simulate only relative product-ion yields, which are scaled to the measured data by a single, global constant. For comparison with scattering experiments measured at a valve temperature $T_{\mathrm{valve}}'$ where no velocity and density profiles were measured, $\rho_{{{\mathrm{H}}_2}}^{T_{\mathrm{valve}}}(z,t)$ was interpolated.

The density $\rho_{{{\mathrm{H}}_2}}^{T_{\mathrm{valve}}}(z,t)$ does not depend on the radial distance from the beam axis, as verified by calculations based on the direct-simulation-Monte-Carlo (DSMC) method~\cite{schullian15a}. However, the geometric truncation of the beam by the skimmer and additional apertures (see Fig.~\ref{fig1}) is explicitly taken into account. 
\subsection{Distribution of product-ion-velocity vectors}
\label{vectors}

The distribution of velocity vectors of the product ions in the laboratory frame depends on the center-of-mass-velocity vector $\vec{v}_{\mathrm{c.m.}}$ of the colliding pair, the collision energy $E_\mathrm{col}$, the differential cross section
$\frac{\mathrm{d}\sigma}{\mathrm{d}E_\mathrm{r}}$ with respect to the recoil energy $E_\mathrm{r}$, and the differential cross sections $\frac{\mathrm{d}\sigma}{\mathrm{d}\theta}$ and $\frac{\mathrm{d}\sigma}{\mathrm{d}\phi}$ with respect to the center-of-mass polar ($\theta$) and azimuthal ($\phi$) angles.

The initial relative velocity of the reactants is obtained from the calculated trajectory $\vec{r}(t)$ of a Rydberg molecule and from the velocity distribution of ground-state H$_2$ molecules $\vec{v}_{\mathrm{H}_2}(\vec{r},t)$ as
\begin{equation}
\vec{v}_{\mathrm{rel,e}}(\vec{r}(t),t)=\dot{\vec{r}}(t) - \vec{v}_{\mathrm{H}_2}(\vec{r}(t),t) \; .
\label{relvelocity}
\end{equation}
The corresponding collision energy is
\begin{equation}
E_{\mathrm{col}}(\vec{r}(t),t)=\frac{1}{2} \, \mu_\mathrm{e} \, |\vec{v}_{\mathrm{rel,e}}(\vec{r}(t),t)|^2,
\label{ecol}
\end{equation}
where $\mu_\mathrm{e}$ is the reduced mass of the reactants.

\begin{figure}
  \centering
\includegraphics[width=0.55 \textwidth]{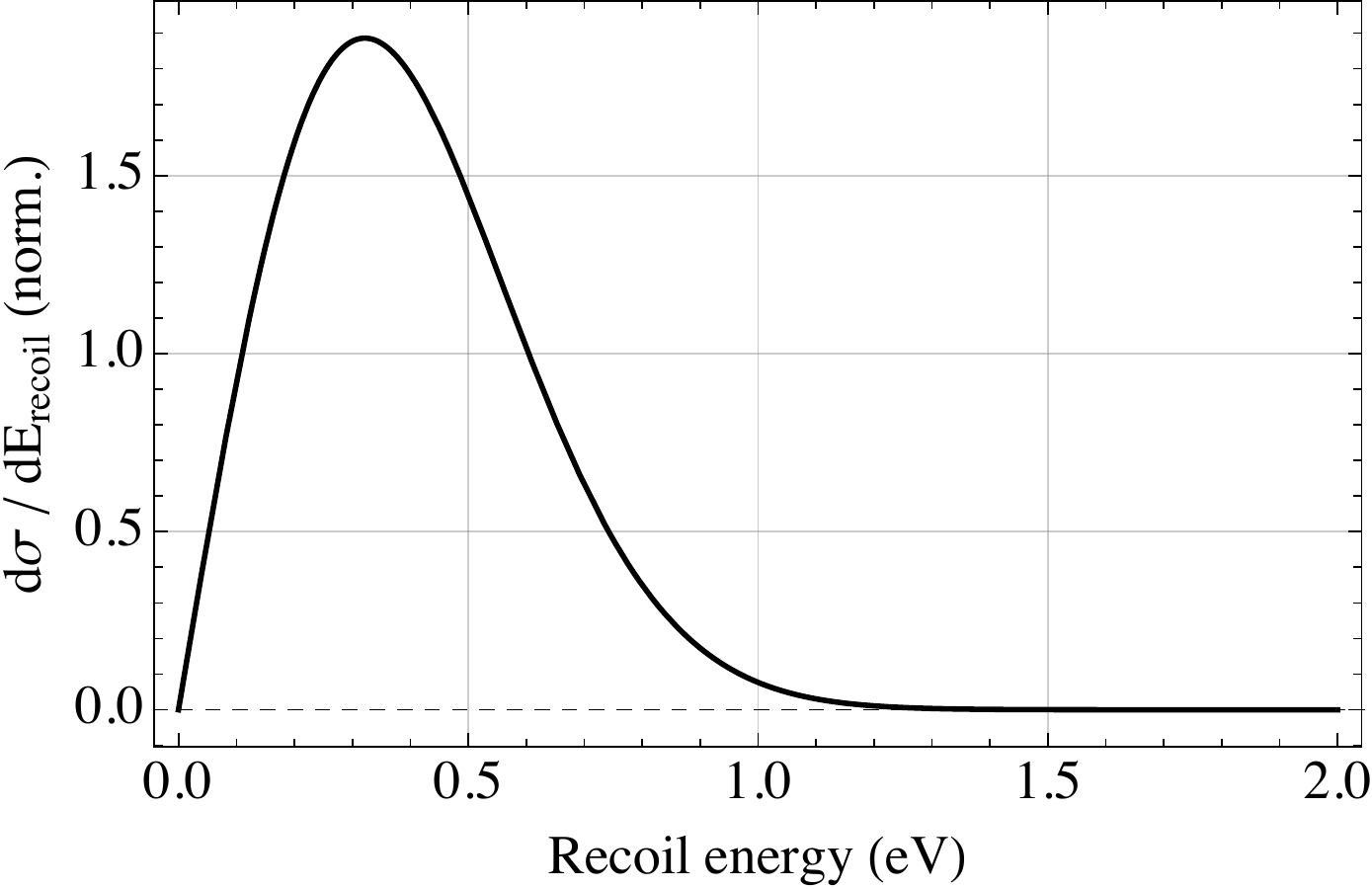}
\caption{Normalized differential cross section $\mathrm{d}\sigma / \mathrm{d}E_{\mathrm{r}}$ of the
${\rm H}_2^+ + {\rm H}_2\rightarrow {\rm H}_3^+ + {\rm H}$ reaction
as a function of product recoil energy $E_{\mathrm{r}}$ used in the simulations.}
\label{fig3}
\end{figure}

The final relative velocity of the reaction products, is given by the recoil energy, as
\begin{equation}
v_{\mathrm{rel,p}}(\vec{r}(t),t)=\sqrt{ \frac{2}{\mu _{\mathrm{p}}}   E_{\mathrm{r}}},
\label{finalvelocityrel}
\end{equation}
where $\mu_{\mathrm{p}}$ is the reduced mass of the reaction products. Previous studies observed that, for low collision energies, about one third of the available energy $E_\mathrm{tot}$, \textit{i.e.}, the sum of $E_\mathrm{col}$ and the exothermicity of the reaction, appears as recoil energy of the products \cite{pollard91}. For the simulation, the recoil energies are randomly sampled from the normalized differential cross section $\frac{\mathrm{d}\sigma}{\mathrm{d}E_\mathrm{r}}$ shown in Fig.~\ref{fig3}. The form of $\frac{\mathrm{d}\sigma}{\mathrm{d}E_\mathrm{r}}$ was modeled using the differential cross section measured by Pollard \textit{et al.} \cite{pollard91} for $E_\mathrm{tot} = 3.25$~eV and scaled linearly to the value of $E_\mathrm{tot} \approx 1.7$~eV in our experiments.

The polar angle $\theta$ of the product-velocity vectors with respect to the initial relative velocity $\vec{v}_{\mathrm{rel,e}}(\vec{r}(t),t)$ is randomly sampled from the normalized differential cross section $\frac{\mathrm{d}\sigma}{\mathrm{d}\theta}$, which was assumed to be uniformly distributed following observations made by Pollard \textit{et al.} \cite{pollard91}. In their studies at low collision energies ($E_\mathrm{col}=1.5$~eV, Fig.~11(a) of Ref.~\cite{pollard91}), H$_3^+$ ions were observed to scatter preferentially into solid angles along the collision axis with only a weak forward-backward asymmetry, which is expected for a strongly exothermic reaction with no internal barrier. The azimuthal angle was also sampled randomly from a uniform distribution.

The velocity vector of the H$_3^+$ product ion in the laboratory frame, taking into account the conservation of momentum, is then given by
\begin{equation}
\vec{v}_{\mathrm{H}_3^+}(\vec{r}(t),t)=\vec{v}_{\mathrm{c.m.}}+\frac{m_{\mathrm{H}}}{m_{\mathrm{H}}+m_{\mathrm{H}_3^+}} \vec{v}_{\mathrm{rel,p}}(\vec{r}(t),t) .
\label{H3velocity}
\end{equation}
A classical trajectory simulation for a H$_3^+$ ion created at $t_{\mathrm{init}}$ and position $\vec{r}(t)$ with velocity vector $\vec{v}_{\mathrm{H}_3^+}(\vec{r}(t),t)$ in the time-of-flight mass spectrometer is then performed in SIMION to test if this ion would be detected in the experiment (item \ref{enum:detection}).

\section{Results}\label{results}

\subsection{Determination of relative reaction cross sections}
\label{cross-sections}

The collision energies can be adjusted in several ways: (i) by changing the temperature of the valve releasing the ground-state H$_2$ beam and thus its velocity (see Fig.~\ref{fig2}d); (ii) by varying the velocity of the Rydberg H$_2$ beam selected by the deflector or by accelerating the Rydberg H$_2$ beam during the deflection \cite{allmendinger14a}; (iii) by modifying the trigger time of the valve releasing the ground-state H$_2$ beam and thus modifying the velocity of the ground-state H$_2$ molecules interacting with the Rydberg H$_2$ molecules in the reaction zone.
In the experiments presented in this article, the collision-energy dependence of the cross section was recorded by monitoring the H$_3^+$ product yield for different temperatures of the valve releasing the ground-state H$_2$ molecules and by systematically varying the trigger time of the valve opening with respect to the laser pulse used to excite the H$_2$ molecules in the second beam to Rydberg states.

\begin{center}
\begin{figure}[ht!]
\includegraphics[width=15 cm]{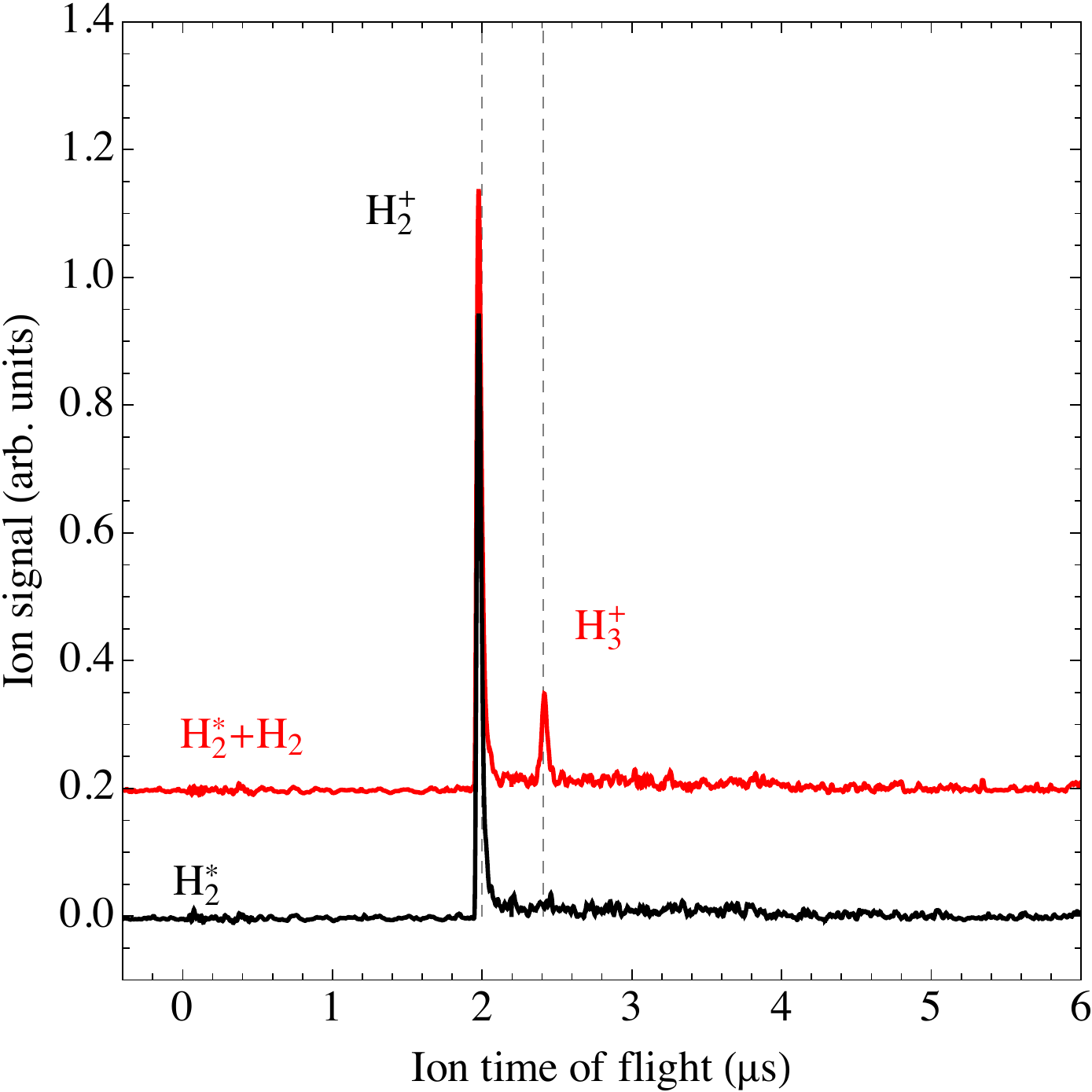}
\caption{Ion time-of-flight spectra recorded in the reaction zone with the ground-state beam turned on (red trace, shifted along vertical axis for clarity) and off (black trace).
}
\label{fig4}
\end{figure}
\end{center}

Fig.~\ref{fig4} compares two ion-time-of-flight spectra recorded by extracting the ions from the reaction zone with a pulsed electric field of 26~V/cm.  The measurement was carried out at a collision energy of $k_{\rm B}\cdot 2.4$~K with H$_2$ Rydberg states initially prepared in the $n=22,k=12,|m|= 3$ Rydberg-Stark state. The valves used to produce the H$_2$ ground-state beam and the Rydberg H$_2$  beam were operated at temperatures of $-133$ and $-170^\circ$C, respectively. To record the red and black traces, which were each obtained by averaging over 50 experimental cycles, the ground-state H$_2$ beam was turned on and off, respectively. In the former case, H$_3^+$ product ions are observed in addition to the H$_2^+$ ions produced by pulsed field ionization of the H$_2$ Rydberg molecules. This H$_2^+$ signal only represents the (small) fraction of the initially prepared $n=22$ Rydberg molecules that underwent a blackbody-radiation-induced transition to Rydberg states with $n>60$, as mentioned in Section~\ref{experiment}. This signal has the same strngth in both traces shown in Fig.~\ref{fig4}, which indicates that only very few of the initially prepared H$_2$ Rydberg molecules undergo a reaction. The assumption of low reaction probability per H$_2$ Rydberg molecule, which is used in the simulations (see Section~\ref{simulations}), is thus justified. This measurement also demonstrates that no significant H$_3^+$ signal is generated from the Rydberg H$_2$ beam alone but that both beam are necessary to observe H$_3^+$ reaction products. 

In reactions with H$_2$ Rydberg molecules prepared in states with principal quantum number $n$ between 22 and 37,  we verified that the number of detected H$_3^+$ ions  does not depend significantly on the amplitude of the pulsed electric extraction field, which was varied between 10 and 60 V/cm. This demonstrates that the H$_3^+$ ions do not result from field-ionization of H$_3^*$ Rydberg molecules, in agreement with the observation of Pratt~\textit{et al.}~\cite{pratt94} that long-lived H$_3^*$ Rydberg molecules do not contribute significantly to the products of the collision. Such molecules would indeed decay by autoionization before the ion-extraction pulse is applied.

\subsection{Comparison of cross sections measured at different $n$ values}
\label{n-dependence}

\begin{center}
\begin{figure}[ht!]
\includegraphics[height=11 cm]{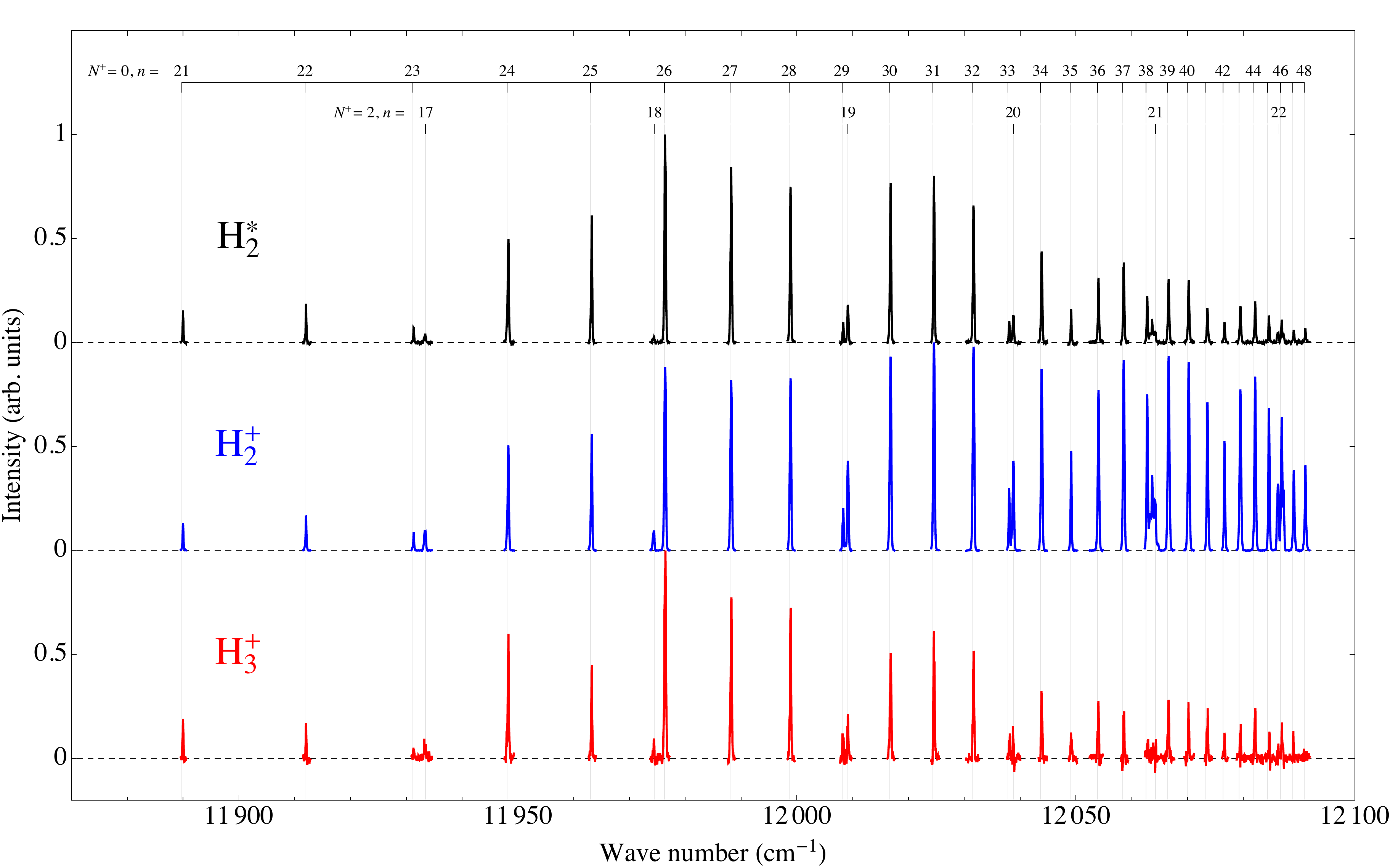}
\caption{
Spectra of the H$_2$ Rydberg series converging on the X$^+$ $^2\Sigma_g^+(v^+=0,N^+=0,2)$ ionization thresholds recorded from the I $^1\Pi_g(v=0,N=0)$ intermediate state. (a) Signal corresponding to the ionization of the Rydberg molecules as they impinge on the MCP detector located beyond the reaction zone (labeled MCP1 in Fig.~\ref{fig1}). (b) H$_2^+$ signal obtained by pulsed field ionization in the reaction zone and extraction of the ions to the MCP detector located at the end of the time-of-flight tube (labeled MCP2 in Fig.~\ref{fig1}). (c)  H$_3^+$ signal resulting from the H$_2^*+{\rm H}_2$ reaction at a collision energy of 90 $\mu$eV ($E_{\rm col}/k_{\rm B}= 1$~K) and extracted to the MCP located at the end of the time-of-flight tube.
}
\label{fig5}
\end{figure}
\end{center}

The H$_3^+$ ion signal obtained from the reaction, though small, is sufficient to record excitation spectra of the Rydberg states of H$_2$. The spectrum displayed in Fig.~\ref{fig5}c was obtained by monitoring the H$_3^+$ ion signal resulting from the H$_2^*$ + H$_2$ reaction at a collision energy of 90~$\mu$eV ($E_{\rm col}/k_{\rm B}= 1$~K). The intensity distribution in this spectrum was used to quantify the effects of a possible dependence of the reaction cross section on the principal quantum number of the excited H$_2$ Rydberg states. To this end, two other excitation spectra of the same series were recorded and are presented in Figs.~\ref{fig5}a and b.

The first of these (Fig.~\ref{fig5}a) was obtained by monitoring the ionization of the H$_2$ Rydberg molecules as these reach the detector MCP1 (see Fig.~\ref{fig1}). When recording this spectrum, the ground-state H$_2$ beam was turned off. The intensity distribution in this spectrum reflects both the excitation probability and the decay of the Rydberg states from the excitation spot (see Fig.~\ref{fig1}) to the detector surface. It is very similar to that observed in Fig.~\ref{fig5}c, with the main difference that the intensities of the transitions to lower (higher) Rydberg states appear slightly weaker (stronger) because of the decay of the Rydberg states during the flight time of the molecules to the detector. Indeed, the decay rate of Rydberg states decreases with increasing $n$ value, which introduces a bias favoring the detection of high Rydberg states in Fig.~\ref{fig5}a. To correct for this effect, additional spectra were recorded using a movable MCP detector (labeled D in Fig.~\ref{fig1}), which enabled us to quantify the $n$ dependence of the decay and to remove its effect on the intensity distributions.

\begin{center}
\begin{figure}[ht!]
\includegraphics[width=15 cm]{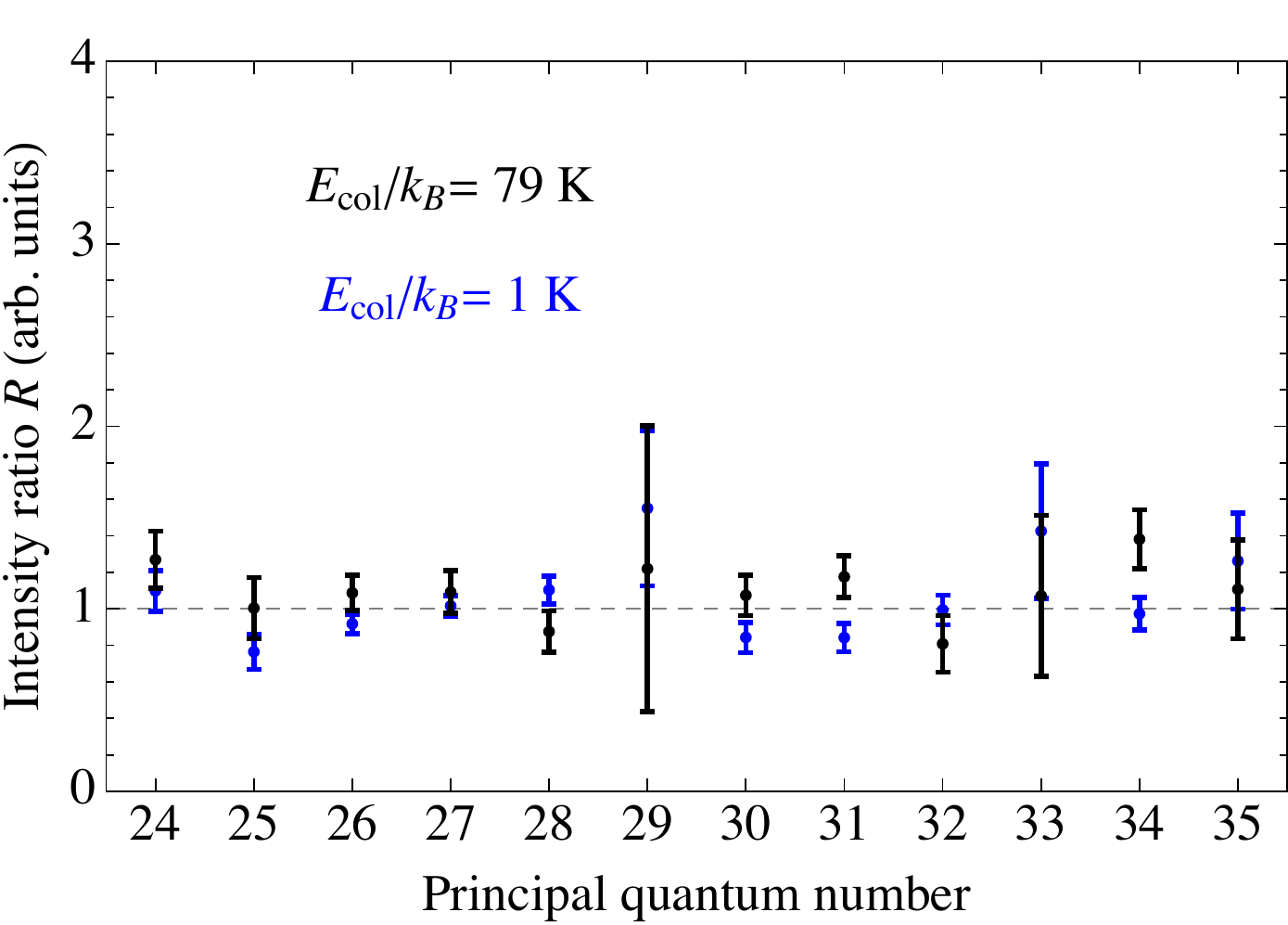}
\caption{Ratio $R$ of the H$_3^+$ signal resulting from the H$_2^*+{\rm H}_2$ reaction to the H$_2^*$ signal detected on MCP1 (signal in Fig.~\ref{fig5}(a)) corrected for the measured $n$-dependent lifetimes. The blue and black data points were obtained for valve temperatures of $-140^\circ$C and $25^\circ$C, respectively, corresponding to $E_{\rm col}/k_{\rm B}=1$~K and 79~K, respectively. The error bars represent one standard deviation. Dashed line: Linear function with slope of $0.003\pm0.013$ obtained by linear regression of the experimental data points (see text for details).
}
\label{fig6}
\end{figure}
\end{center}

The ratio of the intensity of the H$_3^+$ signal from Fig.~\ref{fig5}c to the intensity of the H$_2^*$ signal from Fig.~\ref{fig5}a, but corrected for the decay, is presented in Fig.~\ref{fig6} for $n$ values between 24 and 35 for two representative sets of measurements carried out at collision energies $E_{\rm col}/k_{\rm B}= 1$~K (black dots and error bars) and 79~K (blue dots and error bars), corresponding to the lowest and highest collision energies studied in this work. A linear regression of these data yielded a slope of $0.003\pm 0.013$, which indicates that the cross-sections for the reaction forming H$_3^+$ do not reveal any $n$ dependence within the precision limit of our measurements, i.e., within 1.3~\%, in the investigated range of Rydberg states ($n=24$-35). Over this range of $n$ values, the classical radius of the Rydberg-electron orbit, which scales as $n^2$, increases by a factor of more than 2, and the Rydberg-electron density close to the H$_2^+$ ion core, which scales as $n^{-4}$ in Rydberg-Stark states, decreases by a factor of about 5. We therefore conclude that the influence of the Rydberg electron on the reaction is negligible and that the cross-sections extracted from our measurements faithfully represent the cross sections of the ${\rm H}_2^+ + {\rm H}_2\rightarrow {\rm H}_3^+ + {\rm H}$ reaction within the 1.3~\% precision limit mentioned above.

The excitation spectrum displayed in Fig.~\ref{fig5}b was recorded by monitoring the field ionization signal of the H$_2$ Rydberg molecules in the reaction zone with a pulsed field of 26~V/cm under the same experimental conditions as used to record the spectrum depicted in Fig.~\ref{fig5}c. Such a field only efficiently ionizes Rydberg states with $n$ values beyond 60. The mechanism leading to the observation of the H$_2^+$ signal observed in this spectrum is blackbody-radiation-induced transitions to $n>60$ Rydberg states taking place in the 50~$\mu$s interval between photoexcitation at the entrance of the deflector and the detection in the reaction zone. The gradual increase of the intensity of the field-ionization signal with $n$ in this spectrum is fully consistent with this mechanism and can be modeled with the procedure described in detail in Ref.~\cite{seiler16a}. The reason for showing this spectrum in Fig.~\ref{fig5}~is that it enables us to rule out any significant contribution to the ${\rm H}_3^+$ product yield from reactions of the H$_2^+$ ions produced by pulsed field ionization. Would such reactions significantly contribute to the H$_3^+$ product yield, the H$_3^+$ signal would reveal an increase with $n$, in contrast with the experimental observations. Analysis of the experimental data leads to the conclusion that no significant contribution to the H$_3^+$ signal is produced from such reactions.

\subsection{Collision-energy dependence of the relative cross sections in the range $E_{\rm col}/k_{\rm B}=5$-60~K} \label{T-dependence}

\begin{center}
\begin{figure}[ht!]
\includegraphics[width=15 cm]{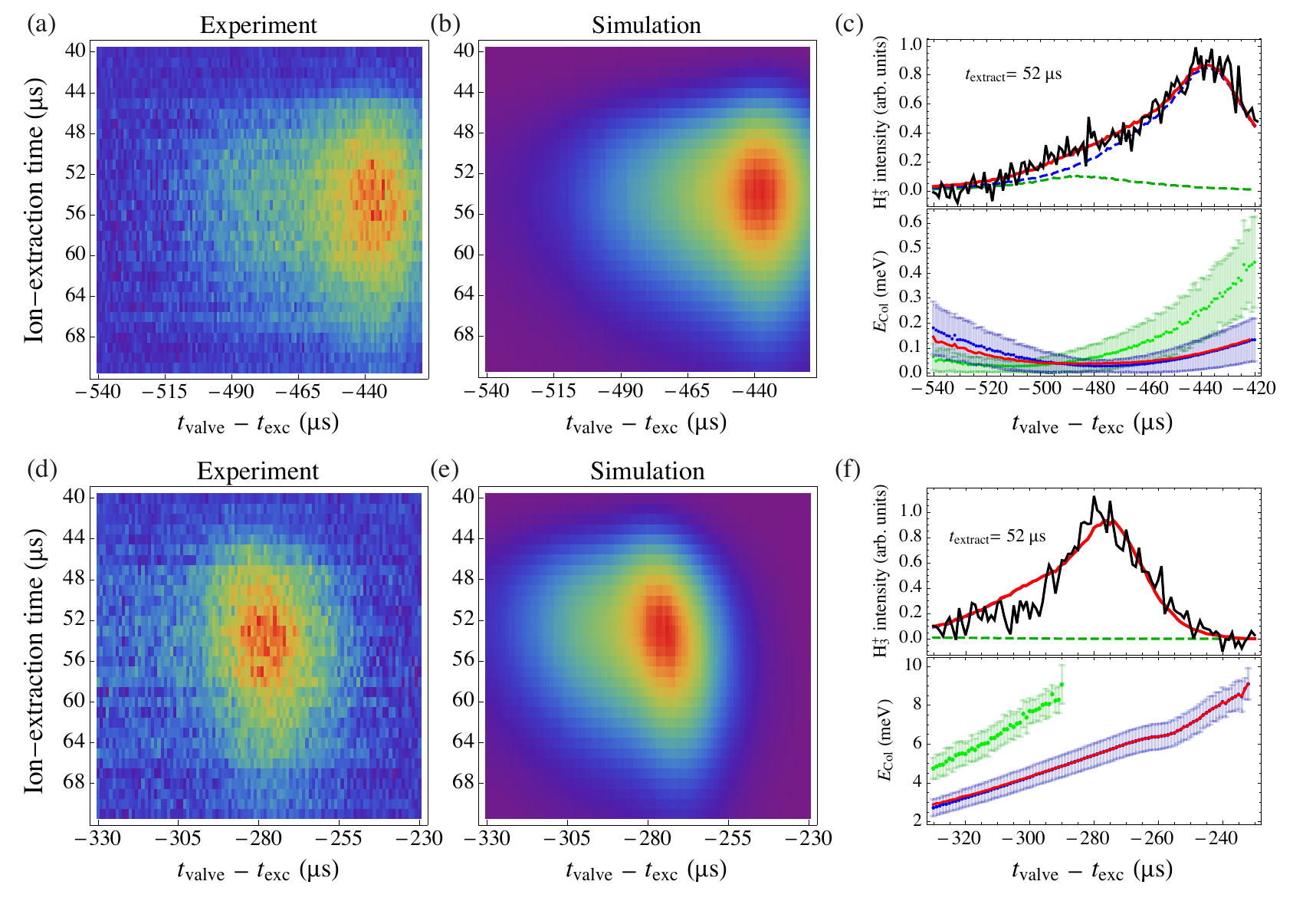}
\caption{(a, b, d, and e) False-color images of the H$_3^+$ signal resulting from the H$_2^*+{\rm H}_2$ reaction as a function of the H$_2$-valve switch-on time $t_{\mathrm{valve}}$ (horizontal axis) and of the time of the ion-extraction pulse (vertical axis) with respect to the Rydberg-excitation laser-pulse trigger time $t_{\mathrm{exc}}$. Panels (a) and (d) present the experimental data recorded at the valve temperatures of (a) -140 $^\circ$C and (d) 25 $^\circ$C. The corresponding simulated images, obtained assuming a uniformly distributed differential cross section $\frac{\mathrm{d}\sigma}{\mathrm{d}\theta}$ are shown in panels (b) and (e), respectively.
(c and f) Upper panel: Cuts through the experimental and simulated images at the ion extraction time of 52~$\mu$s. The black, red, blue, and green traces correspond to the experimental data, their simulations, and the contributions to the simulations from the main pulse and the plunger-rebounce pulse, respectively. Lower Panel: Dependence of the collision energy on the delay between $t_{\mathrm{valve}}$ and $t_{\mathrm{exc}}$. The blue dots with error bars correspond to the main pulse, the green dots with error bars correspond to the plunger-rebounce pulse and the red line to the average of the two energies weighted by their contribution to the detected H$_3^+$ signal.
}
\label{fig7}
\end{figure}
\end{center}

Panels (a) and (d) of Fig.~\ref{fig7} display false-color plots of the H$_3^+$ product yield recorded as a function of the delay time between the opening of the valve releasing the H$_2$ ground-state beam and the ion-extraction pulse (horizontal axis) and of the delay time between the laser pulse used to prepare the H$_2$ Rydberg molecules at the entrance of the deflector and the rising edge of the electric-field pulse used to extract the ions in the reaction zone (vertical axis).

Varying the opening time of the valve releasing the H$_2$ ground-state beam enables us to vary the kinetic energy of the ground-state H$_2$ molecules that react with the Rydberg H$_2$ molecules in the reaction zone, and so to vary the collision energy, as explained in Subsection~\ref{velocity}. The experiments were carried out at a ground-state H$_2$ valve temperature of $-140^\circ$C (panel (a), mean velocity of 1680~m/s) and 25$^\circ$C (panel (d), mean velocity of 2670~m/s) and a mean velocity of the Rydberg H$_2$ beam of 1540~m/s, corresponding to mean collision energies $E_{\rm col}/k_{\rm B}$ of 1~K and 79~K, respectively.
The H$_3^+$ signal intensity along the vertical axis first increases with increasing delay of the ion-extraction pulse because a progressively larger fraction of the ground-state H$_2$ beam has interacted with the Rydberg H$_2$ molecules by the time the pulse is applied. The H$_3^+$ ion signal then returns to zero when the extraction pulse is applied after all H$_3^+$ ions have moved out of the reaction zone.

The variation of H$_3^+$ signal intensity along the horizontal axis primarily reflects the density of ground-state H$_2$ molecules because the rate coefficient $k_{\rm eff}=\langle v \sigma(v) \rangle$ of the reaction is independent of the collision energy, at least within the validity limits of the classical-Langevin-capture model. This is illustrated by the cuts through the images made at an ion-extraction time of 52~$\mu$s  depicted as black traces in the upper panels of Figs.~\ref{fig7}(c) and (f). These two panels also show simulations of the relative H$_3^+$ product yield assuming a pure classical-Langevin-capture behavior as blue, green, and red traces for the contributions of the main pulse, of the plunger-rebounce pulse and their sum, respectively.  

The lower panels of Figs.~\ref{fig7}(c) and (f) illustrate the dependence of the collision energy on the delay between the opening time of the ground-state H$_2$ valve ($t_{\rm valve}$) and the laser pulse used to generate the H$_2$ Rydberg states ($t_{\rm exc}$) for the main gas pulse (blue dots with error bars) and for the minor pulse caused by the plunger rebounce (green dots with error bars). The H$_3^+$ signal at each value of $t_{\rm exc}-t_{\rm valve}$ is proportional to the rate coefficient (see Eq.~(\ref{rate})). Because all other parameters determining the H$_3^+$ signal are known up to a global scaling factor from the measurements and simulations of velocity and density distributions (see Sections~\ref{experiment} and \ref{simulations}), the energy dependence of the rate coefficient can be reconstructed by matching the simulated H$_3^+$ signal to the measured signal at each point of the data set.  The data set involves a large number of measurements such as those depicted in Figs.~\ref{fig7}(a) and (d), recorded for many different temperatures of the ground-state H$_2$ valve. Our values of the rate coefficient thus emerge from a highly redundant data set and are robust against statistical signal fluctuations collected at particular sets of experimental conditions. The extracted rate coefficients faithfully reproduce all recorded data, as illustrated with the two examples presented in Fig.~\ref{fig7}. The known collision energy allows us to convert the measured rate coefficients into cross sections (see Fig.~\ref{fig10} below).

The images presented in Fig.~\ref{fig7} are also sensitive to the assumed distribution of H$_3^+$ product-ion velocity vectors (see Subsection~\ref{vectors}, in particular Eq.~(\ref{H3velocity})) because the H$_3^+$ product ions emitted in the directions parallel or anti-parallel to the collision vector (and thus predominantly moving in a direction parallel or anti-parallel to the merged-beams-propagation direction) have different detection probabilities as a result of purely geometrical considerations. Indeed,  H$_3^+$ ions emitted in the merged-beams propagation direction leave the reaction zone sooner than those emitted in the opposite direction because of the center-of-mass motion of the reactants.

\begin{center}
\begin{figure}[ht!]
\includegraphics[width=15 cm]{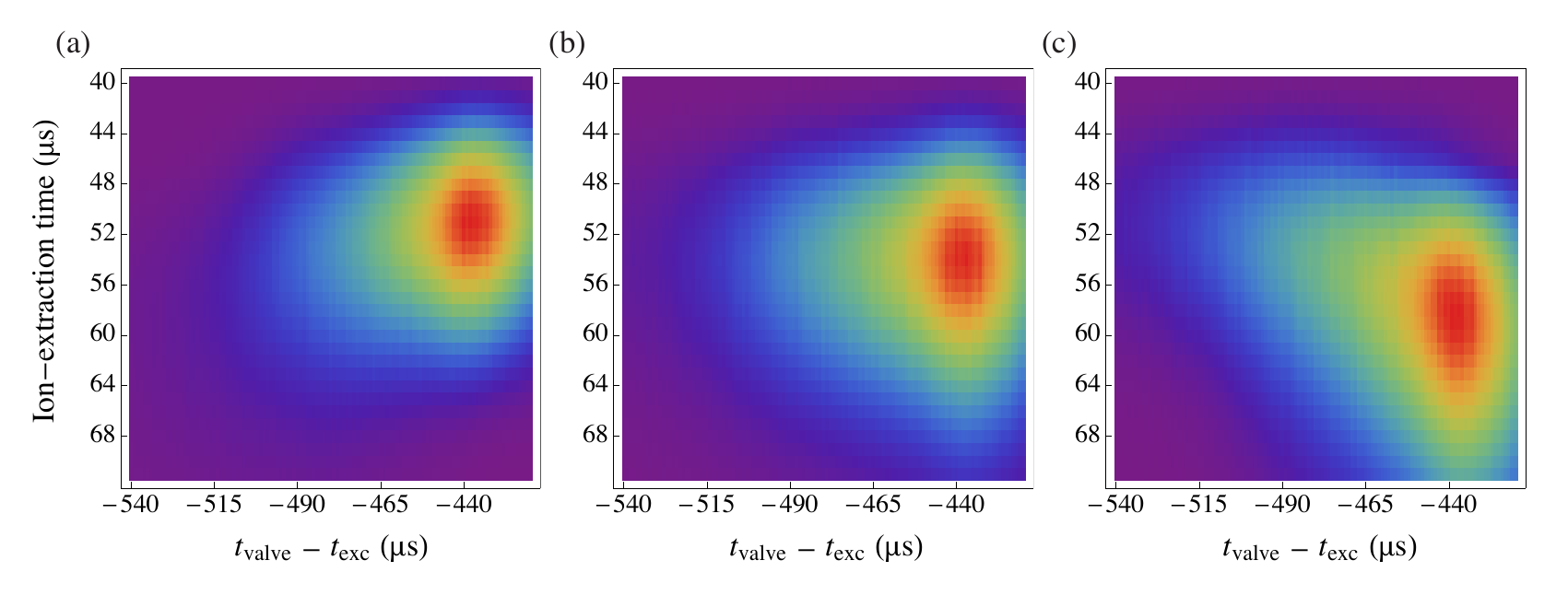}
\caption{Simulations of the experimental results presented in Fig.~\ref{fig7}(a) assuming (a) a maximal forward anisotropy of the product emission, (b) a uniformly distributed differential cross section $\frac{\mathrm{d}\sigma}{\mathrm{d}\theta}$ (already shown in Fig.\ref{fig7}(b)) and (c) a maximal backward anisotropy of the product emission.
}
\label{fig8}
\end{figure}
\end{center}

This aspect is illustrated in more detail in Fig.~\ref{fig8}, which presents three calculations corresponding to the experimental conditions used to record the data presented in Fig.~\ref{fig7}(a), but assuming H$_3^+$ emission parallel to the H$_2^+$ + H$_2$ collision vector (panel (c)), H$_3^+$ emission antiparallel to this vector (panel (a)), and assuming a uniform distribution of the differential cross section $\frac{\mathrm{d}\sigma}{\mathrm{d}\theta}$ (panel (b)) (see Subsection~\ref{vectors}). The experimental data only closely resemble the simulations assuming a uniform $\frac{\mathrm{d}\sigma}{\mathrm{d}\theta}$ distribution, which supports the conclusions drawn by Pollard et al. \cite{pollard91} that the H$_3^+$ product ions are emitted predominantly along the collision axis with only weak (if at all present) backward-forward asymmetry.

\begin{center}
\begin{figure}[ht!]
\includegraphics[width=15 cm]{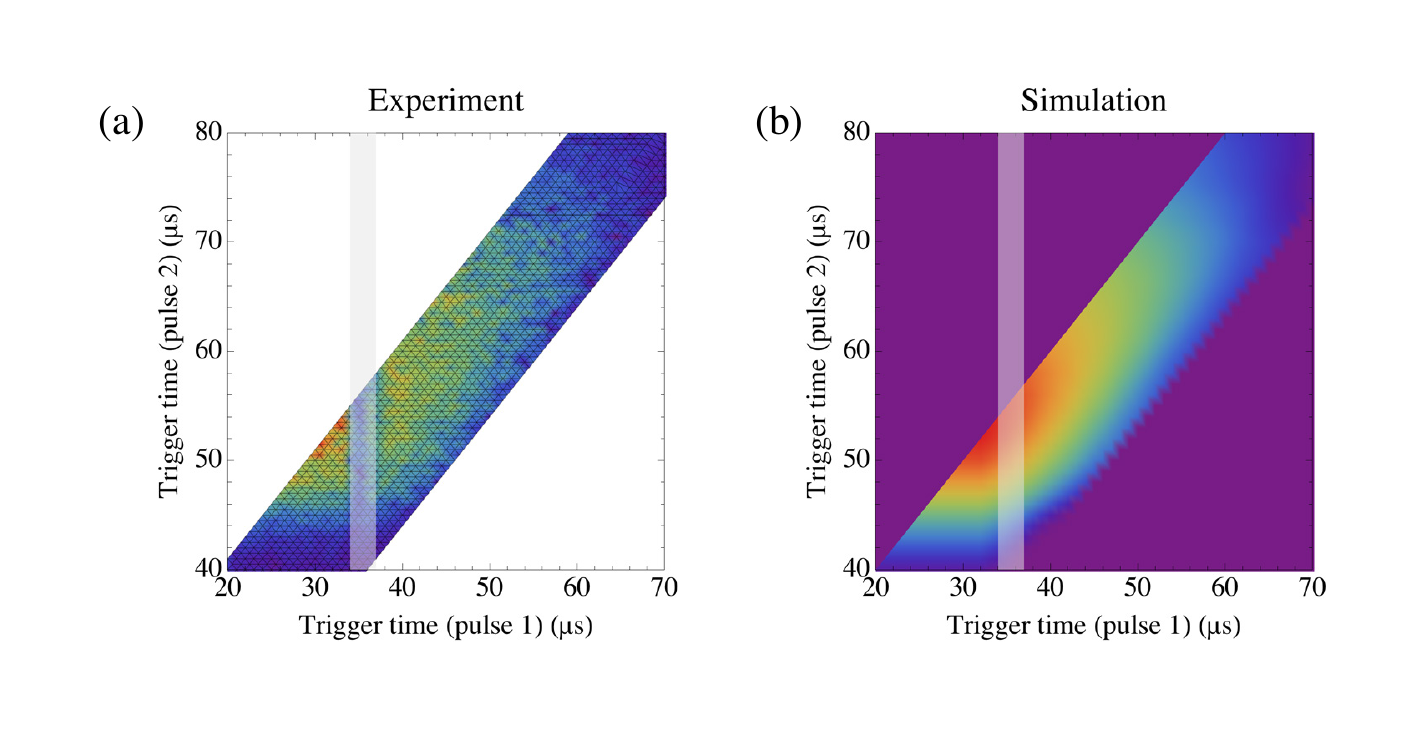}
\caption{Two-extraction-pulse method to reduce the reaction time. Panels (a) and (b) display, respectively, the measured and simulated H$_3^+$ reaction yield extracted by the second extraction pulse as a function of the trigger times of the two extraction pulses. The lower-right data-less triangle corresponds to situations where the second pulse is applied before the first one. In the upper-left data-less triangle, the second pulse is applied more than 20~$\mu$s after the first pulse and does not yield additional information on the reaction.
The area shaded in grey corresponds to the time interval during which all ions are eliminated by the large field generated near the entrance of the reaction zone by the pulsed potential used to generate the first field pulse. See text for additional information.
}
\label{fig9}
\end{figure}
\end{center}

The cross section can also be determined from the two-extraction-pulse experiments described in Subsection~\ref{velocity}. The outcome of experiments in which the trigger time of the first (discrimination) and second (detection) pulses were varied independently are presented in Fig.~\ref{fig9} where they are compared to numerical simulations. The data-less triangle on the lower right side of the figure corresponds to situations where the discrimination pulse is applied after the detection pulse, whereas the data-less triangle on the upper-left side of the figure corresponds to situations where the second pulse is applied too long after the first pulse so that either no signal is observed any more or the energy resolution is degraded. The electric potentials used to generate the discrimination and detection pulses produce very high fields at the entrance of the reaction zone, where the electrodes of the stack are only spaced by a few millimeters (see Fig.~\ref{fig1}). The large electric fields caused in that area by the discrimination pulse effectively prevent the detection of product ions generated in the corresponding time window, which is indicated by the grey-shaded bars in panels (a) and (b) of Fig.~\ref{fig9}. 
The procedure to determine the cross sections from this data set is the same as described for Fig.~\ref{fig7}. The excellent agreement between simulation and experimental results supports the validity of the assumed differential cross sections (see Section~\ref{simulations}).

Fig.~\ref{fig10} presents the energy-dependent cross sections extracted from our measurements as well as cross sections reported in earlier studies in the form of a double-logarithmic plot. The black data points are the absolute integral cross sections obtained by Glenewinkel-Meyer and Gerlich \cite{glenewinkel-meyer97}, the dashed black line represents a pure Langevin-capture behavior, and the green line corresponds to the classical-molecular-dynamics-with-quantum-transitions (MDQT) calculations of Sanz-Sanz {\it et al.} \cite{sanz-sanz15}, respectively. The blue and red dots represent our measurements using $n=31$ and $n=22$ H$_2$ Rydberg states, respectively, and scaled (arbitrarily) so as to match the cross section calculated by Sanz-Sanz {\it et al.} \cite{sanz-sanz15} (their equation (27)) at the collision energy $E_{\rm col}/k_{\rm B}=30$~K.
The vertical error bars indicate the standard deviations of the cross sections and the horizontal bars represent the energy resolution of our measurements.

Although our data are compatible with the data of Sanz-Sanz {\it et al.} \cite{sanz-sanz15} within the resolution and uncertainty limits of our measurement, our data set indicates a slightly steeper slope which is almost identical to the slope of the data of Glenewinkel-Meyer and Gerlich in the range of collision energies between 5 and 10~meV. 

Our two data sets obtained for different $n$ values are consistent, which further demonstrates that the Rydberg electron does not significantly modify the cross sections. The general behavior revealed by the cross sections presented in Fig.~\ref{fig10} indicates that the classical-Langevin-capture model can be used in good approximation when extrapolating the cross-sections of the ${\rm H}_2^+ + {\rm H}_2\rightarrow {\rm H}_3^+ + {\rm H}$ to low collision energies, over the entire range of energies relevant for astrochemistry - down to $E_{\rm col}/k_{\rm B}=5$~K, at least for the lowest rovibrational levels of H$_2$ and H$_2^+$. Deviations caused by quantum-mechanical effects are only expected to occur at lower energies than those  investigated here \cite{dashevskaya05}.

\section{Conclusions}

We have presented a new approach to study ion-molecule reactions at low collision energies, which relies on the equivalence between ion-molecule reactions in free-space and within the orbit of a highly excited Rydberg electron. This equivalence, which we have experimentally verified to high precision with Rydberg-Stark states of principal quantum number in the range 22-40, enables one to avoid the adverse effects of stray electric fields on the collision-energy resolution of cross-section measurements. Indeed, the Rydberg electron effectively shields the reaction from such fields without significantly affecting its outcome.

To reach collision energies below $E_{\rm col}/k_{\rm B}=10$~K, we have exploited a Rydberg-Stark surface-electrode deflector to merge a beam of ground-state molecules with a beam of molecules excited to high Rydberg states. We chose the ${\rm H}_2^+ + {\rm H}_2\rightarrow {\rm H}_3^+ + {\rm H}$ reaction to illustrate the main features of this new approach because of the importance of this reaction in astrophysics and because recent calculations of integral cross sections have been reported for this reaction \cite{sanz-sanz15} with which our results can be compared. Our results indicate that the classical-Langevin-capture model for ion-induced-dipole interactions represents an excellent approximation in the range of  collision energies investigated, down to $E_{\rm col}/k_{\rm B}=5$~K,
as pointed out earlier \cite{dashevskaya05,sanz-sanz15}.

To extract reliable relative cross sections over a broad range of collision energies, care had to be taken to carefully measure the velocity distributions of both beams in the reaction zone and to fully simulate all aspects of the reaction, from the preparation of the molecules in the supersonic beam to the detection of the product ions. The numerical simulation of the experiment enabled us to verify the observation made earlier and at higher collision energies by Pollard {\it et al.} \cite{pollard91} that the product distribution is peaked in the backward and forward directions with no strong forward-backward asymmetry. It also enabled us to assess the collision-energy-resolution limit of the current experiment ($k_{\rm B}\cdot 0.35$~K), which originates from the translational temperature of the H$_2$ Rydberg molecule released by the deflector. Our simulations indicate that reducing the depth
of the moving electric trap used to deflect the Rydberg molecules should permit an improvement of the resolution to about 100 mK. Experiments are currently underway to reach collision energies below 1~K and to study the deviations from the classical-Langevin-capture model predicted by Dashevskaya {\it et al.} \cite{dashevskaya05}.

The method presented and illustrated in this article has the advantage of being applicable to a broad range of ion-molecule reactions. It also permits the full selection of the quantum state of the ion through excitation of Rydberg states belonging to series converging on specific rovibrational levels of the ion core. The experiments presented here were carried out with H$_2^+$ ions in their X $^2\Sigma_g^+\ (v^+=0,N^+=0)$ ground state and with a statistical mixture of para and ortho H$_2$ molecules in their X $^1\Sigma_g^+\ (v=0,N=0)$ (25\%) and $(v=0,N=1)$ (75\%) ground state. Experiments with a pure beam of para H$_2$ molecules as well as with HD, D$_2$, HD$^+$ and D$_2^+$ are envisaged as next steps in our investigations. Reactions involving H$^+$ and He$^+$, which we can control with our surface-electrode decelerator \cite{hogan12,allmendinger13}, could be investigated without substantial modifications of our apparatus and experimental procedure.

\section*{Acknowledgments}

We thank Prof. J. Troe and Prof. E. Nikitin for their encouragements and fruitful discussions. Several results presented here were first presented in November 2015 at a Bunsen Discussion Meeting held in G{\"o}ttingen in honor of Professor J{\"u}rgen Troe, to whom it is a great pleasure to dedicate this article.
This work is supported financially by the Swiss National Science Foundation under project Nr. 200020-149216 and the NCCR QSIT.

\newpage
\section*{Figures}

\begin{center}
\begin{figure}[ht!]
\includegraphics[width=15 cm]{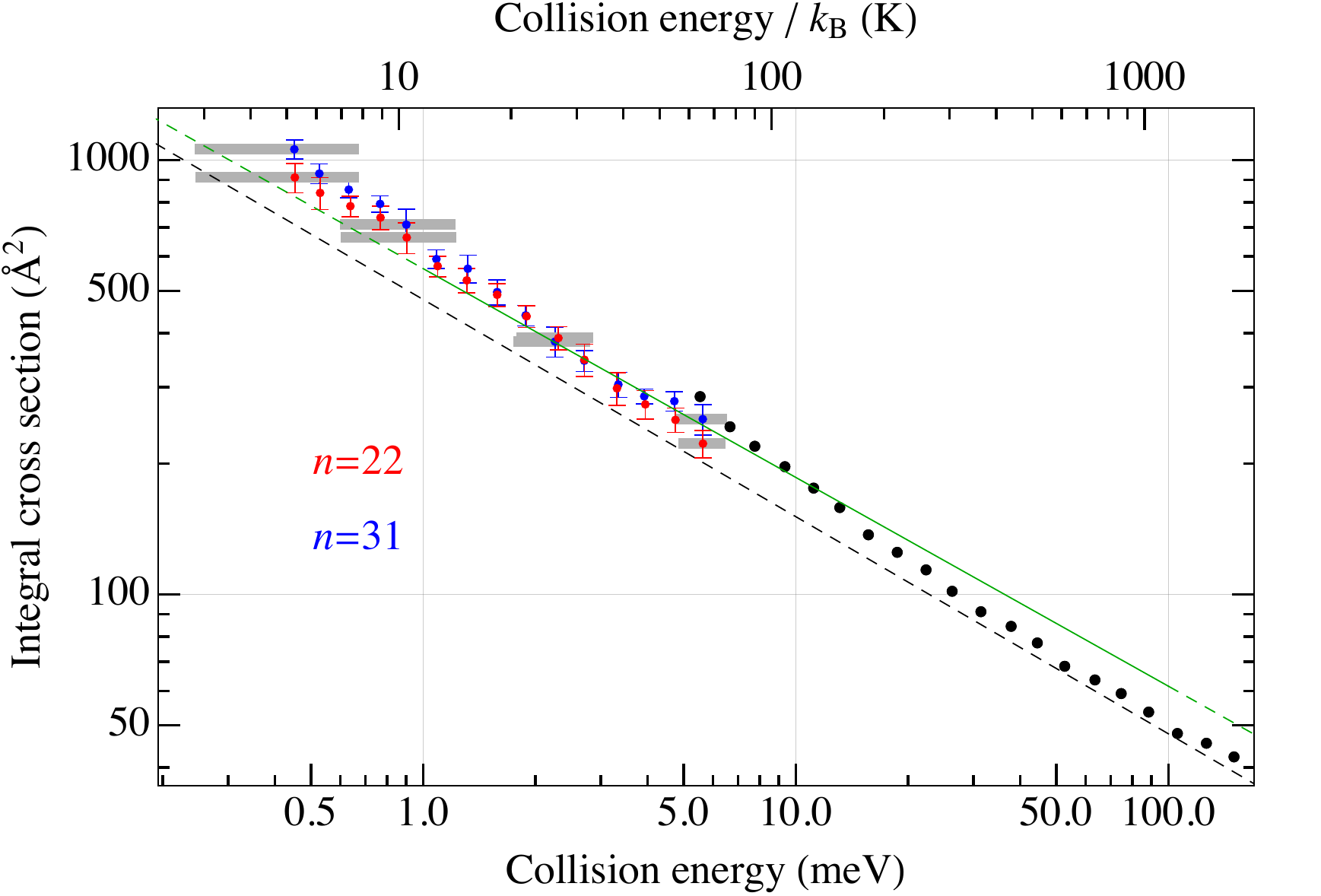}
\caption{Double-logarithmic plot of the absolute reaction cross section of the ${\rm H}_2^+ + {\rm H}_2\rightarrow {\rm H}_3^+ + {\rm H}$ reaction. The black data points are the absolute cross-section measurements of Glenewinkel-Meyer and Gerlich \cite{glenewinkel-meyer97}. The dashed black line and the green line represent a pure Langevin-capture behavior and the classical-molecular-dynamics-with-quantum-transitions calculations of Sanz-Sanz {\it et al.} \cite{sanz-sanz15}, respectively. The blue and red dots represent our measurements using $n=31$ and $n=22$ H$_2$ Rydberg states, respectively, and scaled so as to match the cross section calculated by Sanz-Sanz {\it et al.} \cite{sanz-sanz15} (their equation (27)) at the collision energy $E_{\rm col}/k_{\rm B}=30$~K. The vertical error bars indicate the standard deviation of the measured cross section and the horizontal bars represents the energy resolution of our measurements for representative points.
}
\label{fig10}
\end{figure}
\end{center}

\end{document}